\documentclass[twocolumn,showkeys,pre]{revtex4}

\usepackage{graphicx}%
\usepackage{dcolumn}
\usepackage{amsmath}
\usepackage{latexsym}
\usepackage{longtable}

\begin {document}

\title
{On the origin of the deviation from the first order kinetics in
inactivation of microbial cells by pulsed electric fields }
\author
{N. I. Lebovka$^{1,2}$, E. Vorobiev$^{1}$} \affiliation {$^1$
Departement de Genie Chimique, Universite de Technologie de
Compiegne, Centre de Recherche de Royallieu, B.P. 20529-60205
Compiègne Cedex, France\\ $^2$ Biocolloid Chemistry Institute
named after F. D. Ovcharenko, NASU, bulv. Vernadskogo, 42,
03142,Kyiv, Ukraine}
\begin{abstract}
A computer model was developed for estimation of the kinetics of
microbial inactivation by pulsed electric field. The model is
based on the electroporation theory of individual membrane damage,
where spherical cell geometry and distribution of cell sizes are
assumed. The variation of microbial cell sizes was assumed to
follow a statistical probability distribution of the Gaussian
type. Surviving kinetics was approximated by Weibull equation. The
dependencies of two Weibull parameters (shape \textit{n} and time
$\tau $, respectively) versus electric field intensity E and width
of cell diameters distribution was studied.

\end{abstract}

\maketitle

\section{Introduction}
\label{INTRO}

Pulsed electric fields (PEF) processing is a promising method of
food preservation. Many investigators have shown the effectiveness
of PEF application for killing bacteria in liquid foods
(Barbosa-Canovas et al., 1998; Barsotti \& Cheftel, 1998; Wouters
\& Smelt, 1997). However, there still exist a considerable gap in
understanding the inactivation mechanism. The important problem is
to elucidate how kinetics of killing depends on the type of
bacteria and treatment protocol (electric field strength
\textit{E}, form of pulse, pulse duration t$_{i}$, total time of
treatment).

\begin{figure}
\begin{center}
\includegraphics[width=7.0cm]{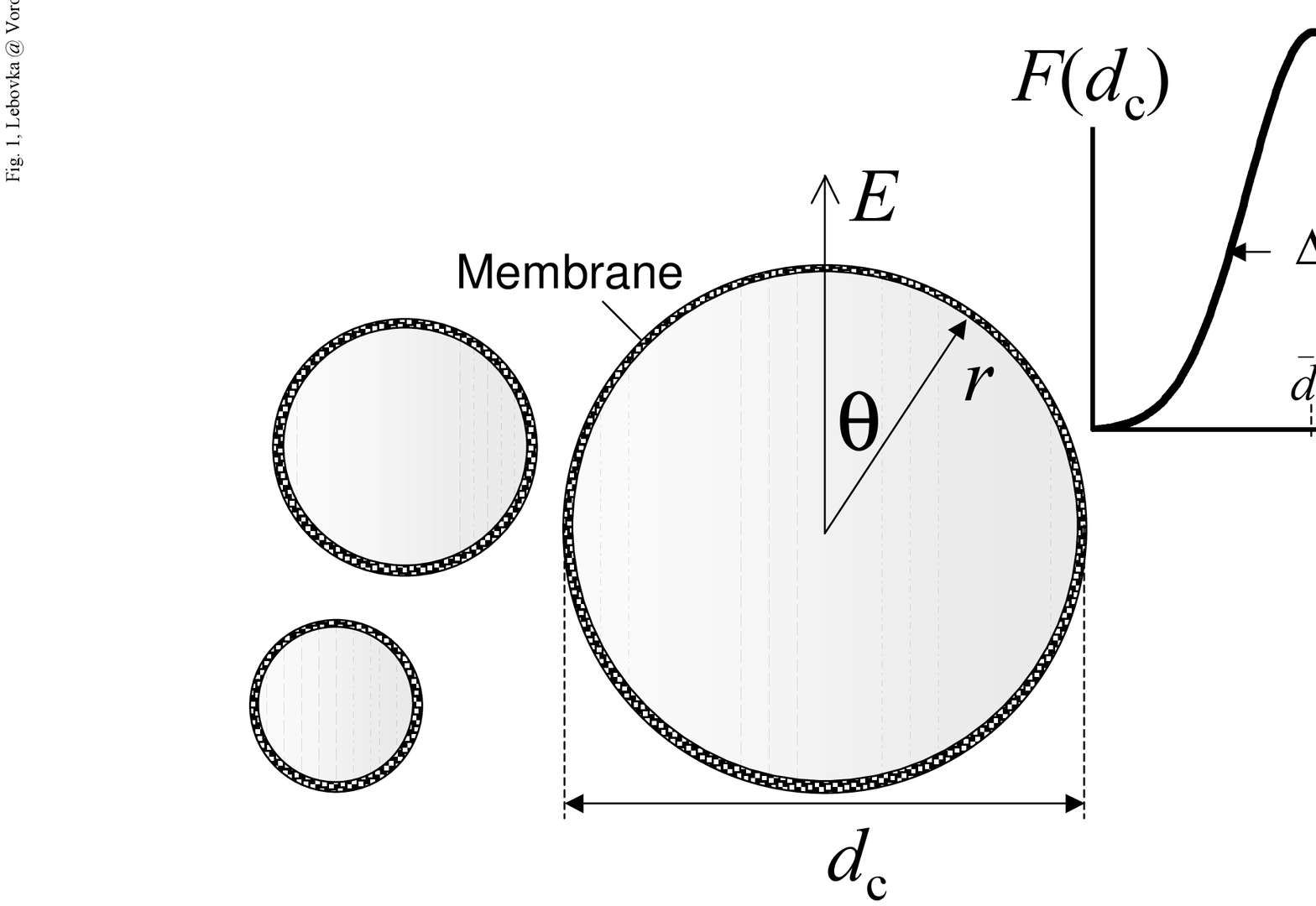}
\end{center}
\caption{Spherical microbial cells in external field \textit{E}.
Cell diameters \textit{d}$_{c}$ are assumed to follow a
distribution function of the Gaussian type
\textit{F}(\textit{d}$_{c}$). Here $\theta $ is an angle between
electric field direction $\vec {E}$ and a radius vector $\vec {r}$
at the surface of membrane.} \label{f01}
\end{figure}

The phenomenon of PEF-inactivation of microbial cells is related
to selective damage of biological membrane. Electrical
conductivity of the membrane $\sigma $ is very low. The reported
values are of order of $\sigma \approx 10^{ - 6} - 10^{ - 7}\Omega
^{ - 1}m^{ - 1}$(Kotnik et al., 1998). Therefore, the highest drop
of potential occurs on the membranes. The transmembrane potential
of a spherical cell $u_{m} $ depends on the angle $\theta $
between the external field $E$ direction and the radius-vector
\textit{r} on the membrane surface, where potential is to be
determined (Fig. 1). This potential may be determined using the
well-known Schwan's equation (Schwan, 1957),
\begin{equation}
u_{m} = 0.75fd_{c} Ecos\theta ,\label{Eq1}
\end{equation}
\noindent where $d_{c} $ is the cell diameter, and $f$ is a
parameter depending on electrophysical and dimensional properties
of the membrane, cell and surrounding media. In dilute suspension
of cells, parameter $f$ is close to 1 (Kotnik et al., 1998).

So, the cell transmembrane potentials $u_{m} $ in suspension of
cells depend on the cell diameter $d_{c} $ and angle $\theta$ and
for the single cell value of $u_{m} $ is maximal at cell poles and
decreases to zero at $\theta = \pm \pi /2$. That is why the
membrane damage probability is maximal at membrane poles and that
bigger sized microbial cells are killed before smaller ones.
Microbial cells always show a variety of shapes and dimensions
(Bergey, 1986). Their size may vary depending on their age, the
nutrients in the growth medium, release mechanisms of microbial
particles etc. (Harding, 1975; Reponen et al., 1992). So,
microbial cells killing probability can also change from cell to
cell.

After PEF treatment during time $t$ in the electric field $E$ the
surviving fraction $S( {t,E} )$ is defined as the ratio of the
number of undamaged microbial cells to the total number of
microbial cells (Barbosa-Canovas et al., 1998). If all the cells
are spherical and are of the same size, then their damage may be
considered as statistically independent events, and the time
dependence $S( {t,E} )$ can be approximated by the first-order
kinetic equation:
\begin{equation} S( {t,E} ) = exp( {
- t/\tau ( {E} )} ),\label{Eq2}
\end{equation}
\noindent where $\tau ( {E} )$ is a time parameter that
corresponds to the effective inactivation time of cells in the
external electric field \textit{E.}

Unfortunately, in most cases, the simplest approximation of the
first-order kinetics is not applicable for description of the
microbial inactivation experiments in pulsed electric fields
(Barbosa-Canovas et al., 1998). Hulsheger et al. (1983) proposed
an empirical equation of type
\begin{equation} S( {t,E}) =
({t/t_{c}})^{ - ({E - E_{c}})/k},\label{Eq3}
\end{equation}
\noindent where $t_{c} $ and $E_{c} $ are the threshold treatment
time and electric field intensity, and \textit{k} is an empirical
parameter. Although this equation is very popular, it has no
theoretical justification.

The models widely used presently for describing the survival
curves are Fermi, log-log and log-logistic models (Barbosa-Canovas
et al., 1998; Peleg, 1996; Alvarez et al., 2000) but they are also
of an empirical nature. Specifically, the Weibull distribution may
be a useful generalization that that includes exponential first
order kinetics as a special case (Peleg, 1995, 1999)
\begin{equation}
S( {t,E} ) = exp( { - ( {t/\tau( {E} )} )^{n( {E} )}}
),\label{Eq4}
\end{equation}
\noindent where $\tau( {E} )$ is a time parameter and $n( {E}
 )$ is a shape parameter. In the case when $n ( {E}
 )$=1, Eq. (4) reduces to Eq. (2).

The time parameter $\tau  ( {E}  )$ in Weibull distribution
accounts for the effective inactivation time, and shape parameter
$n ( {E}  )$ accounts for the concavity of a survival curve (van
Boekel, 2002). Weibull distribution was applied for fitting
experimental PEF inactivation data (Alvarez et al., 2002), but
physical meaning of the obtained parameters $\tau  ( {E}  )$ and
$n ( {E}  )$ was not elucidated yet.

A possible deviation from the first order PEF inactivation
kinetics may be caused by existence of a variety of microbial
shapes and dimensions. The purpose of this paper is to analyse how
the form of the survivor curves can reflect existence of a
distribution of cell diameters.

\section{Computational model and details of calculations}

The Monte Carlo technique was used for simulation of PEF
inactivation kinetics of microbial cells. Initial number of
microbial cell in suspension was put as N$_{o}$=10$^{7}$. A
Gaussian law distribution function of cell diameters was assumed
(Fig.1)
\begin{equation}
F( {d_{c}}) = \frac{{1}}{{\sqrt {2\pi}  \Delta} }exp( { - \frac{{(
{d_{c} - \bar {d}_{c}} )^{2}}}{{2\Delta ^{2}}}} ), \label{Eq5}
\end{equation}
\noindent where $\bar {d}_{c} $ and $\Delta $ represent the
average diameter and the standard deviation, respectively.

\begin{figure}
\begin{center}
\includegraphics[width=7.0cm]{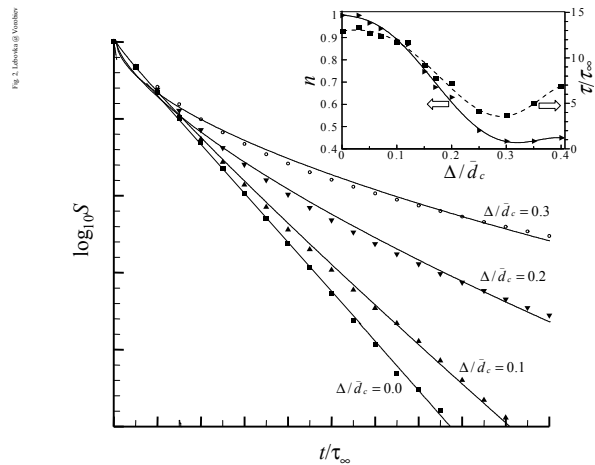}
\end{center}
\caption{Calculated survivor curves $S ( {t/\tau _{\infty} }
 )$. Insert shows the shape $n$ and relative time $\tau
$/$\tau \infty $ parameters of Weibull distribution versus the
relative width of cell diameter distribution $\Delta /\bar {d}_{c}
$. All the calculations were made for $E^{\ast}  = 10$. }
\label{f02}
\end{figure}

An arbitrary microbial cell was chosen in suspension for a given
time $t$. Then, a random point on the membrane surface was chosen
by generating of cos$\theta $ value randomly in the intervals
$-1$-$+1$. The lifetime of a membrane $\tau $ on the surface of a
cell depends on its diameter $d_{c}$, angle $\theta $, and
intensity of external field $E$. It was found on the basis of the
transient aqueous pore model (Weaver \& Chizmadzhev, 1996), that:
\begin{equation}
\tau ( {\theta ,d_{c} ,E} ) = \tau _{\infty} exp\frac{{\pi \omega
^{2}/kT\gamma} }{{1 + ( {u_{m} ( {\theta ,d_{c} ,E} )/u_{o}}
)^{2}}},
 \label{Eq6}
\end{equation}
\noindent where $u_{m}(d_{c},\theta $) was calculated from Eq.
(1). Here, $\tau _{\infty}  $ is the parameter ($\tau \to \tau
_{\infty}  $ in the limit of very high electric fields), $\omega $
and $\gamma $ are the line and surface tensions of membrane,
respectively, $k$ is the Boltzmann constant, $T$ is the absolute
temperature, $u_{o} = \sqrt {2\gamma / ( {C_{m}  ( {\varepsilon
_{w} /\varepsilon _{m} - 1}
 )}  )} $ is the voltage parameter (the dimension of
$u_{o} $ is Volts), $C_{m} $ is the specific capacitance of a
membrane, $\varepsilon _{w} ,\varepsilon _{m} $ are the relative
dielectric permittivities of the aqueous phase and of the
membrane, respectively.

The probability of the chosen cell damage was approximated by the
first-order kinetic equation as $exp ( { - t/\tau  ( {\theta
,d_{c} ,E}  )}  )$. This procedure was repeated for all the cells
in the suspension. Then, the number of killed cells was enumerated
for the given time t, surviving fraction $S ( {t,E}  )$ was
calculated, time was increased by a time step and procedure was
repeated from beginning.

In this work, the voltage scale parameter was estimated as $u_{o}
\approx 0.17$V from data obtained by Lebedeva (1987) for the
general lipid membranes ($\omega \approx 1.69\ast 10^{ - 11}$N,
$\gamma \approx 2\ast 10^{ - 3}$ N/m, $\varepsilon _{w} \approx
80$, $\varepsilon _{m} \approx 2$, $C_{m} \approx 3.5\ast 10^{ -
3}$F/m$^{2}$ at $T = 298$K). The time scale parameter was put as
$\tau _{\infty}  \approx 3.7\ast 10^{ - 7}$s (Lebedeva, 1987).
Dimensionless reduced field intensity was defined as $E^{\ast}  =
E/E_{o} $, where $E_{o} = u_{o} / ( {0.75fd_{c}}   )$ was
estimated as $E_{o} \approx 2.27$ kV/cm at $d_{c} \approx 1\mu m$.

\section{Results and discussion}

Figure 2 presents some examples of the calculated survivor curves
$S ( {t}  )$ for suspension of cells (symbols) at the given value
of reduced field intensity $E^*=10$ ($E \approx $22.7 kV/cm, at
$d_c=1\mu$m). The first order kinetics law is only observed for
suspensions of identical cells ($\Delta /\bar {d}_{c} = 0$). In
other cases, the noticeable deviations from the first order
kinetics are observed and the more pronounced deviations are
observed with increase of $\Delta /\bar {d}_{c} $.

\begin{figure}
\begin{center}
\includegraphics[width=7.0cm]{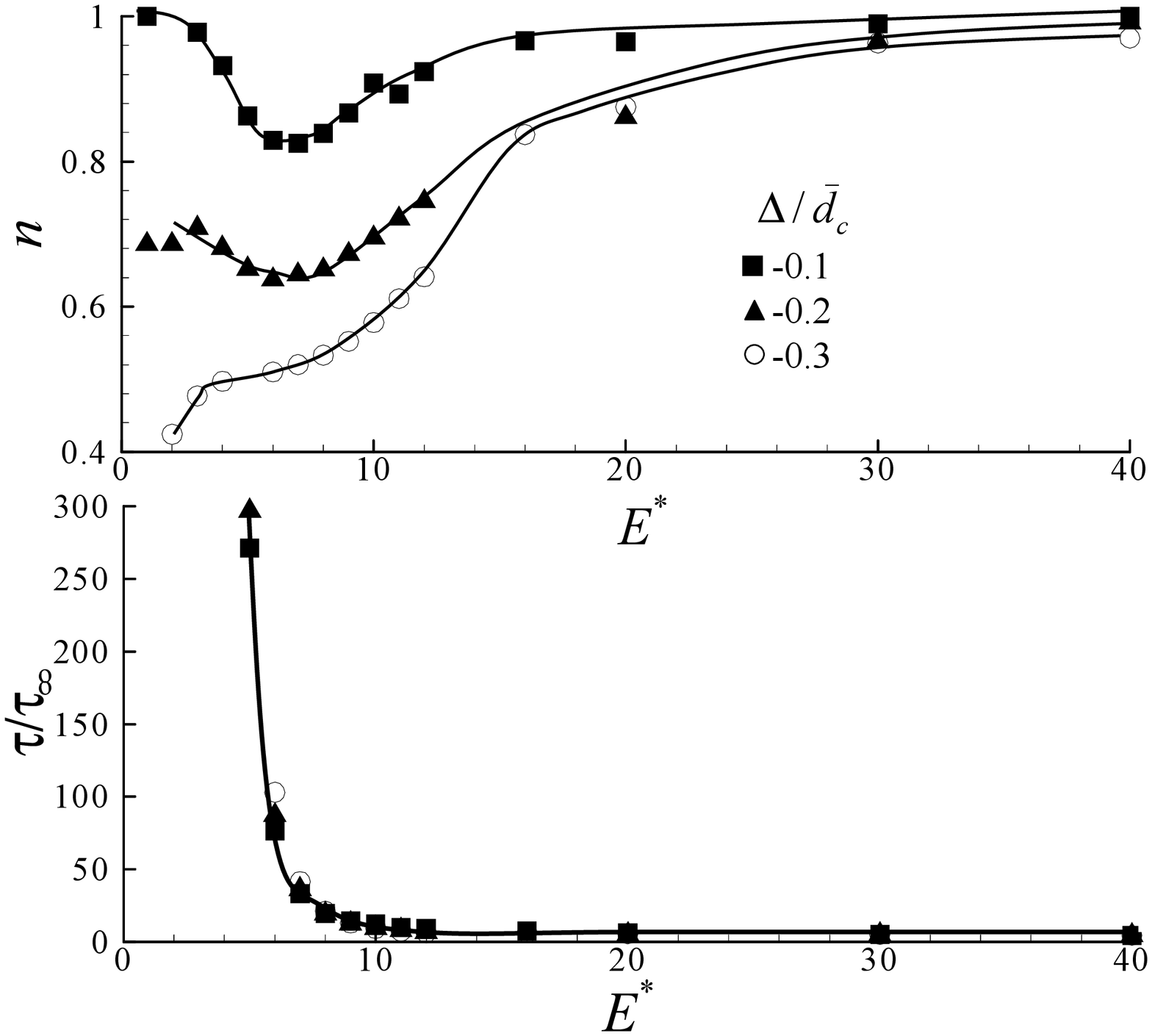}
\end{center}
\caption{Shape $n$ and relative time $\tau $/$\tau \infty
$\textit{ }parameters of Weibull distribution versus reduced
electric field intensity $E^{\ast}  = E ( {0.75fd_{c} /u_{o}}
 )$ for suspension of cells with diameter distribution for
different values of relative width of cells diameter distribution
$\Delta /\bar {d}_{c} $. The simulated data were fitted with the
Weibull equation within time interval of $0<t/\tau_\infty<200$.}
\label{f03}
\end{figure}

The solid lines drawn through the symbols are the best fit to data
simulated using the Weibull function (Eq. 4). The Weibull law
seems to be very appropriate for approximation of calculated
survival curves (in all the cases the correlation coefficient
$\rho $ was higher than 0.993). Insert to Fig.2 shows the shape
$n$ and relative time $\tau/\tau_\infty $ parameters of Weibull
distribution versus relative width of cell diameter distribution
$\Delta /\bar {d}_{c} $. In these estimations the fitting was done
within the time interval $1<t/\tau_\infty <200$. At $\tau
_{\infty} \approx 3.7\ast 10^{ - 7}$s (Lebedeva, 1987) this time
interval corresponds to $0<t<74\mu$s. Both the shape parameter $n$
and the relative time $\tau/\tau_\infty $ parameter initially
decrease with increases of the standard deviation $\Delta $. Then,
a small elevation of these values is observed, which can be
explained by the distortion of the Gaussian distribution at higher
values of $\Delta /\bar {d}_{c} $. But this model always gives
only upward concavity, i.e. $n<1$.

\begin{figure}
\begin{center}
\includegraphics[width=7.0cm]{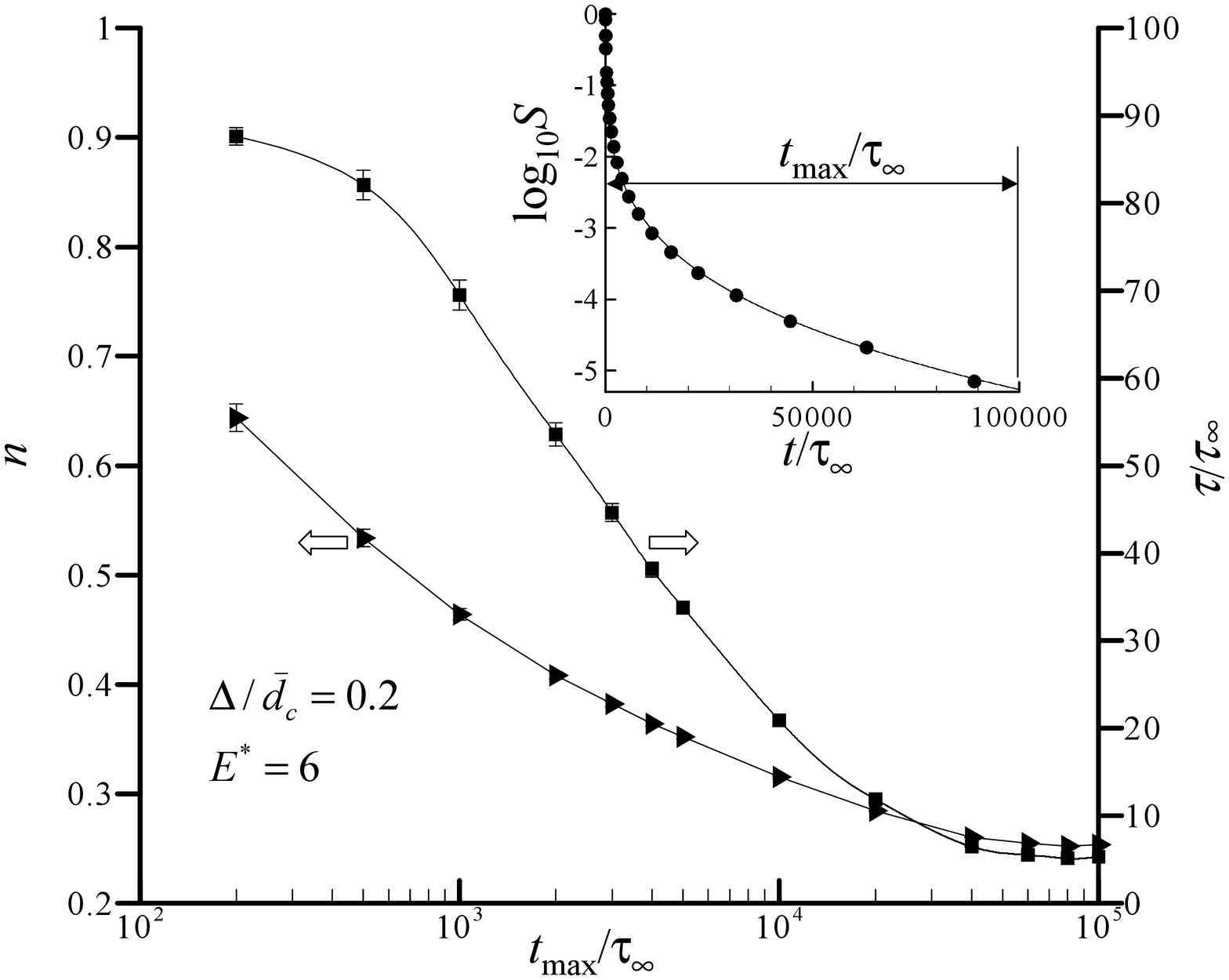}
\end{center}
\caption{Shape parameter $n$ and relative time parameter $\tau
$/$\tau_\infty $ versus time interval $t_{max}/\tau_\infty $.
Calculations were done at reduced electric field intensity of
$E^{\ast} = E ( {0.75fd_{c} /u_{o}}) = 6$ and relative width of
cell diameter distribution $\Delta /\bar {d}_{c} = 0.2$. Dash line
shows the time $t_{max}/\tau_\infty  =200$ used for calculation of
data presented in Fig.1 and Fig. 2. The insert shows calculated
survivor curve $S ( {t/\tau_{\infty} }  )$ for this particular
case. The solid line is the best fit to the simulated data
(symbols) with the Weibull equation in the time interval
$t/\tau_\infty <10^{5}$. } \label{f04}
\end{figure}

Parameters $n$ and $\tau /\tau_\infty $ are also very sensitive to
the value of electric field intensity $E$ (Fig. 3). Dependencies
of $n$ versus $E^*$ are rather complex, but in all cases parameter
$n$ increases with increase of the distribution width $\Delta
/\bar {d}_{c} $. Relative time parameter $\tau/\tau_\infty$
decreases considerably with field intensity $E$ increase, but it
is practically independent of $\Delta /\bar {d}_{c} $. So, it is
possible to conclude that upward concavity of survivor curve is
rather sensitive both to the field intensity and to the
variability of microbial diameter distribution, but the effective
inactivation time $\tau(E)$ is insensitive to the variations in
cell diameters.

The numerically estimated shape $n$ and relative time $\tau
$/$\tau_\infty$ parameters are rather sensitive to the time
interval of Monte Carlo data fitting with Weibull function Eq.
(4). Fig. 4 shows a typical example of $n$ and $\tau $/$\tau_
\infty$ versus $t_{max}$/$\tau_\infty$ dependencies for fitting of
the same survival curve when upper bound of the time interval
$1<t/\tau_\infty<t_{max}/\tau_\infty$ is varied. In fact,
$t_{max}/\tau_\infty$ is a relative total time of treatment. In
all cases, the apparent consistency between Monte Carlo data and
Weibull function with adjusted parameters $n$ and $\tau $/$\tau_
\infty$ was rather good, and the correlation coefficients $\rho$
lied in the interval 0.993-0.998. But at the same time, parameter
$n$ and $\tau/\tau_\infty$ are sensitive to the upper cutting
boundary $t_{max}/\tau_\infty$, and this fact reflects existence
of an intrinsic inconsistency between unknown survival function
and Weibull function.

\section{Conclusion}

The discussed illustrative examples show that geometry of the
survival curve is very sensitive to the distribution of cell
diameters. The Weibull function seems to be appropriate for
approximation of the calculated survival curves. The parameters of
this function $\tau$ and $n$ are rather sensitive to the width of
distribution of the cells diameters, electric field intensity and
total time of treatment. We would note, that the proposed model is
based on several restrictive assumptions. The survival kinetics
may reflect many intrinsic details of the real microbial cells.
For explanation of the survival curves, experimentally observed
for PEF-inactivated population, it is necessary to introduce into
the model the experimentally determined distribution functions of
cell diameters. It is also desirable to use in calculations more
realistic law of an individual membrane damage, based on
experimentally estimated data for the given bacterial population.
The possible effects of sub-lethal damage, when bacterial damage
needs some critical destructive exposure, also were not considered
in this model. A deviation from the first-order kinetics may be
also influenced by others factors, such as existence of bacterial
geometry anisotropy and distribution of bacterial orientations.
So, in future it is seems to be important to find correlations
between variations in factors, influencing bacterial geometry,
dimension distribution function, details of membrane damage and
parameters of bacterial inactivation kinetics. Such work should be
done in order to improve practically important PEF-treatment
regimes for reaching a desirable value of microbial inactivation.

\section*{Acknowledgements}
The authors are indebted to the anonymous referee for helpful
comments on the manuscript and valuable suggestions. The authors
would like to thank the "Pole Regional Genie des Procedes"
(Picardie, France) for providing the financial support. Authors
also thank Dr. N.S. Pivovarova and Dr. A.B. Jemai for their help
with preparation of the manuscript.

\section*{References}

Alvarez, I., Raso, J., Palop, A., Sala, F. J., 2000. Influence of
different factors on the inactivation of Salmonella senftenberg by
pulsed electric fields. International Journal of Food Microbiology
55, 143-146.

Alvarez, I., Pagan, R, Raso, J, Condon, S., 2002. Pulsed electric
field inactivation of Listeria monocytogenes described by the
Weibull distribution. In: Cano, M. P., Morante P. (Eds.), EMERTEC
2002, Symposium on Emerging Technologies for the Food Industry.
Madrid, Spain 11-13 March 2002, p. 116.

Barbosa-Canovas, G.V., Góngora-Nieto, M.M., Pothakamury, U.R.,
Swanson, B.G., 1998. Preservation of foods with pulsed electric
fields. Academic Press, London.

Barsotti, L., Cheftel, J.C., 1998. Traitement des aliments par
champs electriques pulses. Science des Aliments 18, 584-601.

Bergey, L., 1986. Manual of systematic bacteriology. Williams and
Wilkins, Baltimore.

Harding, H., 1975. Effect of pH and sucrose concentration on
conidium size and septation in four \textit{Bipolaris} species.
Canadian Journal of Botany 53, 1457-1464.

Hulsheger, H., Potel, J., Niemann, E.G., 1983. Electric field
effects on bacteria and yeast cells. Radiat. Environ. Biophys. 22,
149-162.

Kotnik, T., Miklavcic D., Slivnik, T., 1998. Time course of
transmembrane voltage induced by time-varying electric fields: a
method for theoretical analysis and its application.
Bioelectrochemistry and Bioenergetics 45, 3-16.

Lebedeva, N.E., 1987. Electric breakdown of bilayer lipid
membranes at short times of voltage effect. Biologicheskiye
Membrany, 4 , 994-998 (in Russian).

Peleg, M., 1995. A model of microbial survival after exposure to
pulsed electric fields. Journal of the Science of Food and
Agriculture 67, 93-99.

Peleg, M., 1996. Evaluation of the Fermi equation as a model of
doze response curves. Applied Microbiology and Biotechnology 46,
303-306.

Peleg, M., 1999. On calculating sterility in thermal and
non-thermal preservation methods. Food Research International 32,
271-278.

Reponen, T., 1995 Aerodynamic diameters and respiratory deposition
estimates of viable fungal particles in mold problem dwellings.
Aerosol Science and Technology 22, 11-23.

Schwan, H. P., 1957. Electrical properties of tissue and cell
suspensions. In: Lawrence, J. H., Tobias, A. (Eds.), Advances in
biological and medical physics, vol. 5. Academic Press, New York,
pp. 147-209.

van Boekel, M.A.J.S., 2002. On the use of the Weibull
model to describe thermal inactivation of microbial vegetative
cells. International Journal of Food Microbiology 74, 139-159.

Weaver, J.C., Chizmadzhev, Y.A., 1996. Theory of electroporation:
a review. Bioelectrochemistry and Bioenergetics 41, 135-160.

Wouters, P. C., Smelt, J.P.P.M., 1997. Inactivation of
microorganisms with pulsed electric fields: Potential for food
preservation. Food Biotechnology 11, 193-229.

\end{document}